\documentclass[10pt]{article}
\usepackage[pctex32]{graphics}
\usepackage[dvips]{graphicx}

\usepackage{epsfig}
\usepackage{graphicx}
\usepackage{indentfirst}
\usepackage{amsmath}
\usepackage{amsfonts}
\usepackage{amssymb}
\usepackage{hyperref}
 \usepackage{latexsym}
\usepackage{color}

\begin{document}

\begin{center}
{\Large\bf Tolman-Oppenheimer-Volkoff Equations and their implications for the structures of relativistic Stars in $f(T)$ gravity }

\medskip

A. V. Kpadonou$^{(a,b)}$\footnote{e-mail: vkpadonou@gmail.com}, 
M. J. S. Houndjo$^{(a,c)}$\footnote{e-mail:
sthoundjo@yahoo.fr} and  
M. E. Rodrigues$^{(d,e)}$\footnote{e-mail: esialg@gmail.com}

$^a$ \,{\it Institut de Math\'{e}matiques et de Sciences Physiques (IMSP)}\\
{\it 01 BP 613,  Porto-Novo, B\'{e}nin}\\
$^{b}$\,{\it Ecole Normale Sup\'erieure de Natitingou - 
Universit\'e de Natitingou - B\'enin} \\
$^{c}$\,{\it Facult\'e des Sciences et Techniques de Natitingou - 
Universit\'e de Natitingou - B\'enin} \\
$^{d}$ \,{\it  Faculdade de F\'isica, PPGF, 
Universidade Federal do Par\'a, 66075-110, Bel\'em, Par\'a, Brazil.}\\   
$^{e}${\, \it Faculdade de Ci\^encias Exatas e Tecnologia, 
Universidade Federal do Par\'a - Campus Universit\'ario de Abaetetuba, 
CEP 68440-000, Abaetetuba, Par\'a, Brazil }\\
\end{center}
\date{}

\begin{abstract}
We investigate in this paper the structures of neutron and quark stars in 
$f(T)$ theory of gravity where $T$ denotes the torsion scalar. Attention is 
attached to the TOV type equations of this theory and numerical integrations 
of these equations are performed with suitable EoS. We search for the 
deviation of the mass-radius diagrams for power-law and exponential type 
correction from the $TT$ gravity. Our results show that for some values of 
the input parameters appearing in the considered models, $f(T)$ theory 
promotes more the structures of the relativistic stars, in consistency with 
the observational data. 

\end{abstract}

Pacs numbers: 


\section{Introduction}
The current acceleration of the universe is widely accepted through various 
independent observational data, as supernovae Ia \cite{1a3deasuivre}-
\cite{1a3deasuivrefin}, cosmic microwave background radiation 
\cite{4a6deasuivre}-\cite{4a6deasuivrefin}, the large scale structure of the 
universe \cite{7a8deasuivre}-\cite{7a8deasuivrefin}, cosmic shear through 
gravitational weak lensing surveys \cite{5deodintsov} and the Lyman alpha 
forest absorption lines \cite{4deodintsov}. The well known standard equivalent 
theories, the General Relativity (GR) and the Tele-parallel Theory, are the 
first theories used for explaining the acceleration of the universe, including 
the existence of the dark energy as a new component of the universe 
\cite{7a15deodintsov}-\cite{7a15deodintsovfin}.\par

Another way of explaining this acceleration of the universe is modifying a 
standard theory, $GR$ or $TT$. In this theory we are interested to the 
modification of $TT$, getting the so-called $f(T)$ theory of gravity  where 
$T$ denotes the torsion scalar. Instead of the Levi-Civita connection in the 
$GR$, the $TT$ and $f(T)$ are based on the Weitzenbock connection. Note that 
in $f(T)$  theory of gravity, new scalar degrees of freedom appear, 
constraining the consideration of the torsion scalar into dynamics way as 
effective new scalar field, instead of the torsion scalar in $TT$ defined by 
the pressure and the density inside the stars. In the theories based on the 
curvature scalar, the semi-classical approach to quantum gravity is often used 
where the higher and logarithmic terms are included in the action because their 
relevance for the strong field regime in the interior of the relativistic stars 
\cite{22a24deasuivre}-\cite{22a24deasuivrefin}. Neutron stars  probe high 
baryon densities where baryon density in the stellar interior can be of the 
order of magnitude beyond the nuclear saturation density 
$(\rho_c=2.7\times 10^{17})$ k$\mathrm{g}m^{-3}$ \cite{asuivre}. \par
In this paper, inspired by the fact that in dense medium the strong nuclear 
force play a crucial role, we consider the effect of corrections to the $TT$ 
action involving terms of power-law and exponential in $T$ on the observational 
features of the neutron stars and quark stars with suitable equations of state 
(EoS). For the neutron stars, we consider the polytropic  EoS and the SLy EoS, 
while for the quark stars, we assume a simplest EoS in the so-called bag model. 
In this way, Tolman-Oppenheimer-Volkoff Equations are established from the 
generalized field equation in $f(T)$. First we consider diagonal tetrad and 
fall into the constraint where the algebraic action function is reduced to the 
$TT$ one. Since the goal is to introduce correction terms to the $TT$ theory, 
we consider in the second step a non-diagonal tetrad through which the previous 
constraint fall down. According to our results it comes that for some values of 
the input parameters $b_1$, $b_2$, $n$ and $q$, structures of relativistic 
stars, both neutron and quark stars, are able to be found within $f(T)$ theory 
of gravity.\par
The paper is organized as follows. The section $2$ is devoted to the generality 
on $f(T)$ gravity and the generalization of the TOV equations within diagonal 
tetrad fashion. The TOV equations are still developed in the section $3$ but 
within the non-diagonal fashion; then the structures of the relativistic stars, 
both neutron and quark stars, are analyzed through numerical integrations of 
the field TOV equations. Our conclusion and perspectives are presented in the 
section $4$.
\section{Generality on $f(T)$ gravity and generalised TOV equations }

The action $S$ of $f(T)$ gravity, is given by 
\begin{eqnarray}
\label{fla1}
S=\int d^{4}x e\left[\frac{f(T)}{2\kappa^{2}}+\mathcal{L}_{m}\right],
\end{eqnarray}

where $\mathcal{L}_{m}$ is the matter Lagrangian density assumed to depend only 
on the tetrad, and not on its covariant derivatives;  $e$ is the determinant of 
the tetrad and $f$ a generic function depending on the scalar torsion. \par 
The variation of the action (\ref{fla1}) with respect to the tetrad leads to  
\cite{daouda1,daouda2}
\begin{eqnarray}
\label{fla2}
 S^{\;\;\; \nu \rho}_{\mu} \partial_{\rho} T f_{TT} + 
[e^{-1} e^{i}_{\;\; \mu}\partial_{\rho}(e e^{\;\; \alpha}_{i}S^{\;\;\; \nu\lambda}_{\alpha} )
+T^{\alpha}_{\;\;\; \lambda \mu}   S^{\;\;\; \nu \lambda}_{\alpha} ]f_{T}+
\frac{1}{4}\delta^{\nu}_{\mu}f=\frac{\kappa^{2}}{2} \mathcal{T}^{\nu}_{\mu},
\end{eqnarray}
where $\mathcal{T}^{\nu}_{\mu}$ is the energy-momentum tensor. We consider that 
the matter content is an isotropic fluid, such that the corresponding 
energy-momentum tensor reads
\begin{eqnarray}\label{fla3}
(\rho+p_t)u_\mu u^\nu-p_t\delta^\nu_\mu+(p_r-p_t)v_\mu v^\nu,
\end{eqnarray}
with $u^{\mu}$, $v^{\mu}$  the four-velocity and the unit space-like vector in 
the radial direction. The parameters $\rho$, $p_r$ and $p_t$ denote the energy 
density, the pressure in the direction of $v^\mu$ (normal pressure) and $p_t$ 
the pressure orthogonal to $v^{\mu}$ (transversal pressure), respectively. 
In an isotropic case, one has the equality $p_r=p_t$.\par
In order to get solutions describing stellar objects, we assume a spherically 
symmetric metric with two independent functions $\alpha$ and $\beta$ both 
depending on the radial coordinate, as 
\begin{eqnarray}\label{fla4}
ds^2= e^{\alpha(r)}dt^2-e^{\beta(r)}dr^2-r^2(d\theta^2+\sin^2\theta d\phi^2).
\end{eqnarray} 

According to this metric, the field equation can be decoupled as
\begin{eqnarray}
\frac{f}{4}-\left[T-\frac{1}{r^2}-\frac{e^{-\beta}}{r}\left(\alpha'+\beta'
\right)\right]\frac{f_T}{2}&=&4\pi\rho, \label{fla5}\\
\left(T-\frac{1}{r^2}\right)\frac{f_T}{2}-\frac{f}{4}&=&4\pi p_r, \label{fla6}\\
\Bigg[\frac{T}{2}+e^{-\beta}\Bigg(\frac{\alpha''}{2}+\left(\frac{\alpha'}{4}+
\frac{1}{2r}\right)(\alpha'-\beta')\Bigg)\Bigg]\frac{f_T}{2}-\frac{f}{4}&=&4
\pi p_t,\label{fla7}\\
\frac{\cot\theta}{2r^2}T'f_{TT}&=&0,\label{fla8}
\end{eqnarray}
where the prime denotes the derivative with respect to the radial coordinate 
$r$. The torsion scalar reads
\begin{eqnarray}
T=2\frac{e^{-\beta}}{r^2}\left(1+r\alpha'\right).\label{fla9}
\end{eqnarray}

From the conservation Law for the stress tensor, i.e, 
$\nabla_\mu \mathcal{T}^\mu_\nu=0$, one gets
\begin{eqnarray}\label{fla10}
\frac{dp_r}{dr}=-\frac{1}{2}\left(\rho+p_r\right)\frac{d\alpha}{dr}+\frac{2}{r}
\left(p_t-p_r\right).
\end{eqnarray}
In the usual cases, i.e, with an isotropic fluid, where $p_r=p_t=p$, and 
setting $\alpha \rightarrow 2\alpha$, one gets 
$\frac{dp}{dr}=-(\rho+p)\frac{d\alpha}{dr}$.\par 
By taking into account the trace of the field equations, one gets 
\begin{eqnarray}\label{fla11}
\frac{e^{-\beta}}{4r^2}\left[4(e^\beta-3)+4r(\beta'-3\alpha')+r^2(\alpha'
\beta'-2\alpha''-\alpha'^2)\right]f_T+f=4\pi(\rho-p_r-2p_t).
\end{eqnarray}
In order to get the TOV equations, it is suitable to replace the metric function
by the following expression
\begin{eqnarray}
e^{-\beta}=1-\frac{2GM}{r}\label{fla12}
\end{eqnarray}
For the following considerations, it is convenient to adopt dimensional 
variables $M \rightarrow mM_{\odot}$ and $r \rightarrow r_g r$ with 
$r_g=GM_{\odot}$. Then, from (\ref{fla12}) one can extract 
\begin{eqnarray}\label{fla13}
e^{-\beta}=1-\frac{2m}{r}\Longrightarrow \beta'=\frac{2m}{r^2}
\frac{1-\frac{rm'}{r}}{\frac{2m}{r}-1}.
\end{eqnarray} 

Then it is easy to get stellar structures. It is important to note here that  
some asymptotic flatness are required as the radial coordinate evolves, that is
\begin{eqnarray}\label{fla14}
\lim_{r\rightarrow\infty}T(r)=0,\quad\quad\quad \lim_{r\rightarrow\infty} 
m(r)=cste 
\end{eqnarray}

By making use of(\ref{fla9}), (\ref{fla10}) within the condition of isotropy 
$p_r=p_t=p$ and (\ref{fla13}), one gets in terms of the mass, the pressure and 
the energy density, the following TOV equations
\begin{eqnarray}
\left[\frac{2}{r(\rho+p)}\left(1-\frac{2m}{r}\right)\frac{dp}{dr}+\frac{1}{r^2}
\frac{dm}{dr}-\frac{1}{2r^2}+\frac{m}{r^3}\right]f_T+\frac{1}{4}f-4\pi \rho=0   
\label{fla15}  \\
\left[-\frac{\alpha'}{r}+\frac{2\alpha' m}{r^2}-\frac{1}{2r^2}+\frac{2m}{r^3}
\right]f_T+\frac{1}{4}f+4\pi p=0,\label{fla16}\\
\Bigg[-\frac{2}{\rho+p}\left(\frac{1}{r}\frac{dm}{dr}-\frac{6}{r}+
\frac{11m}{r^2}\right)\frac{dp}{dr}+\frac{4}{(\rho+p)^2}\left(\frac{2m}{r}-1
\right)\left(\frac{dp}{dr}\right)^2\nonumber \\ -2\left(\frac{2m}{r}-1\right)
\frac{d}{dr}\left(\frac{1}{\rho+p}\frac{dp}{dr}\right)-\frac{2}{r^2}
\frac{dm}{dr}-\frac{2}{r^2}+\frac{4m}{r^3}\Bigg]f_T+f-4\pi \left(\rho-3p
\right)=0. \label{fla17}
\end{eqnarray}
In order to solve the system (\ref{fla15})-(\ref{fla17}), we need the expression
of the algebraic function $f(T)$, an explicit form of the equation of state 
$p=f(\rho)$, and some initial conditions on the mass $m$ and the pressure. As 
initial conditions for the mass and the pressure one can assume $m(0)=0$ and 
$p(0)=f(\rho_c)$, where $\rho_c=\rho(0)$. From the equation (\ref{fla8}) two 
cases can be distinguished, i.e, $T=cste$ or $f(T)=T-2\Lambda$, where $\Lambda$ 
is view as the cosmological constant. Let us first work with the realistic case,
that is $f(T)=T-2\Lambda$. 

\section{Stars structures through non-diagonal tetrad}

In this section we consider a non-diagonal tetrad, in order to escape the 
constraint leading to the linear form of the algebraic function $f(T)$. 
We chose the non-diagonal tetrad as
\begin{eqnarray}\label{fla19}
\left\{e^{i}_{\;\;\mu}\right\}=
\begin{pmatrix}
e^{\alpha} & 0 & 0 & 0 \\ 
0 & e^{\beta}\sin\theta\cos\phi & r\cos\theta\cos\phi & -r\sin\theta\sin\phi \\ 
0 & e^{\beta}\sin\theta\sin\phi & r\cos\theta\sin\phi & r\sin\theta\cos\phi \\ 
0 & e^{\beta}\cos\theta & -r\sin\theta & 0
\end{pmatrix} 
\end{eqnarray}
The TOV equations in this case read

\begin{eqnarray}
-\frac{1}{r^2}\sqrt{r(r-2m)}\left(\sqrt{\frac{r-2m}{r}-1}\right)T'f_{TT}
\nonumber\\
+\frac{1}{2r^3}\sqrt{\frac{r}{r-2m}}\Bigg[\left(
\alpha'r^2-2m'r-2\alpha'mr+2r-2m\right)\sqrt{\frac{r-2m}{r}}\nonumber\\-
\alpha'r^2+2\alpha'mr-2r+4m\Bigg]f_T+\frac{1}{4}f=4\pi\rho\label{fla20}\\
\frac{1}{2r^3}\sqrt{r(r-2m)}\Bigg[2(\alpha'r+1)\sqrt{\frac{r-2m}{r}}-\alpha'r-2
\Bigg]f_T-\frac{1}{4}f=4\pi p_{r}\label{fla21}\\
\frac{1}{4r^2}\sqrt{r(r-2m)}\Bigg[(\alpha'r+2)\sqrt{\frac{r-2m}{r}-2}\Bigg]T'
f_{TT}-\frac{1}{8r^3}\Bigg[4r(\alpha'r+2)\sqrt{\frac{r-2m}{r}}\nonumber\\
-2\alpha''r^3-\alpha'^2r^3+2\alpha'm'r^2+4\alpha''mr^2+2\alpha'^2mr^2\nonumber\\
-6\alpha'r^2+4m'r+10\alpha'mr-8r+4m\Bigg]f_T-\frac{1}{4}f=4\pi p_t\label{fla22}
\end{eqnarray}
where $p_r$ and $p_t$ mean the radial and tangential pressures respectively. 
In this case, the trace of the equation of motion leads to
\begin{eqnarray}
\left[\frac{1}{r}\left(2\sqrt{1-\frac{2m}{r}}+\alpha'm-2\right)+\frac{4m}{r^2}-
\frac{\alpha'}{2}\right]T'f_{TT}+\Bigg(
\frac{\alpha'm'}{2r}+\frac{\alpha''m}{r}+\frac{\alpha'^2}{2r}-\frac{3\alpha'}{r}
\nonumber\\
\frac{2m'}{r^2}+\frac{11\alpha'm}{2r^2}-\frac{4}{r^2}+\frac{4m}{r^3}-
\frac{\alpha''}{2}-\frac{\alpha'^2}{4}\Bigg)f_T+f=4\pi \left(\rho-p_r-2p_t
\right)\label{fla23}
\end{eqnarray}
In this rebric we will still work in the frame of MIT bag model where the 
simple equation of state for quark matter is (\ref{fla18}). Here, we will 
distinguish two models, the power law and exponential models.
\subsection{Stars structures within power-law and exponential gravities} 
In this subsection we assume the algebraic power-law action function as 
$f(T)=T+b_1T^n$ where $b_1$ is  real constant and $n$ a real number with 
$n\neq 0$ and $n\neq 1$, such that $f_T(T)=1+nb_1T^{n-1}$ and 
$f_{TT}(T)=n(n-1)b_1T^{n-2}$. The exponential action function is assumed as 
$f(T)=b_2e^{qT}$, with ($b_2\neq 0$ and $q\neq 0$) such that 
$f_T(T)=b_2qe^{qT}$ and $f_{TT}(T)=b_2q^2e^{qT}$.

\subsubsection{Neutron Stars}
The structure of neutron stars has been developed in other kind of modified 
gravity, namely the $f(R)$ gravity \cite{11deasuivre}-\cite{18deasuivre}. 
In our case we will explore the observable features such as the mass-radius 
relation of the neutron stars. In order to solve the generalized field equations
it is needed the relation between the energy density and the pressures. We will 
assume for simplicity an homogeneous matter content such that the tangential and
radial pressures are equal, and equations of state will be assumed for the 
neutron stars driving the information of the matter inside the star. Two types 
of equation of state  are considered: the simpler polytropic EoS and a more 
realistic SLy EoS \cite{49deasuivre}.
\begin{center}
1.  {\it Polytropic EoS}
\end{center}
As simpler polytropic equation of state, we assume the following one
\begin{eqnarray}
\zeta=2\xi+5\label{fla24}
\end{eqnarray}
with 
\begin{eqnarray}\label{fla25}
\xi=\log(\rho/\mathrm{g}\,cm^{-3}), \quad\quad \zeta=\log(p/dyn\,cm^{-2})
\end{eqnarray}
By making use of Runge-Kutta method we numerically solve the system formed by 
the equations (\ref{fla20}), (\ref{fla21}) and (\ref{fla25}) through an 
integration from the center to the surface of the star. We assume that the 
density at the center of the star is increased from a non null value $\rho_c$, 
fixed to $\rho_c=2.7\times 10^{17} kg m^{-3}$, called nuclear saturation 
density. The numerical results are presented and the deviation from the $TT$ 
theory of gravity can be seen for large values, both positive and negative,  of 
the parameters $b_1$ and $b_2$ for the power-law gravity and exponential 
gravity, respectively. The evolution the mass-radius diagram for neutron stars 
within polytropic EoS are represented in Fig $1$ for power-law correction and 
Fig $2$ for exponential correction to the $TT$ of gravity.


\begin{center}
\begin{figure}[t]
\begin{minipage}[t]{0.3\linewidth}
\includegraphics[width=\linewidth]{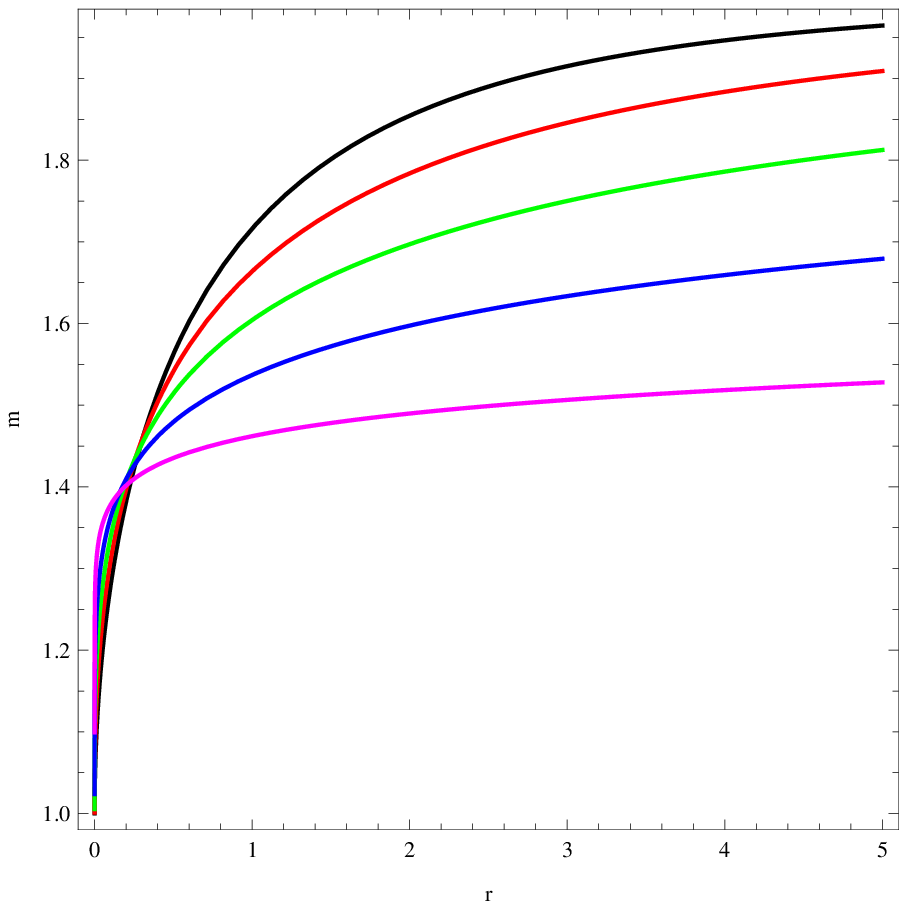}
\end{minipage} \hfill 
\begin{minipage}[t]{0.3\linewidth}
\includegraphics[width=\linewidth]{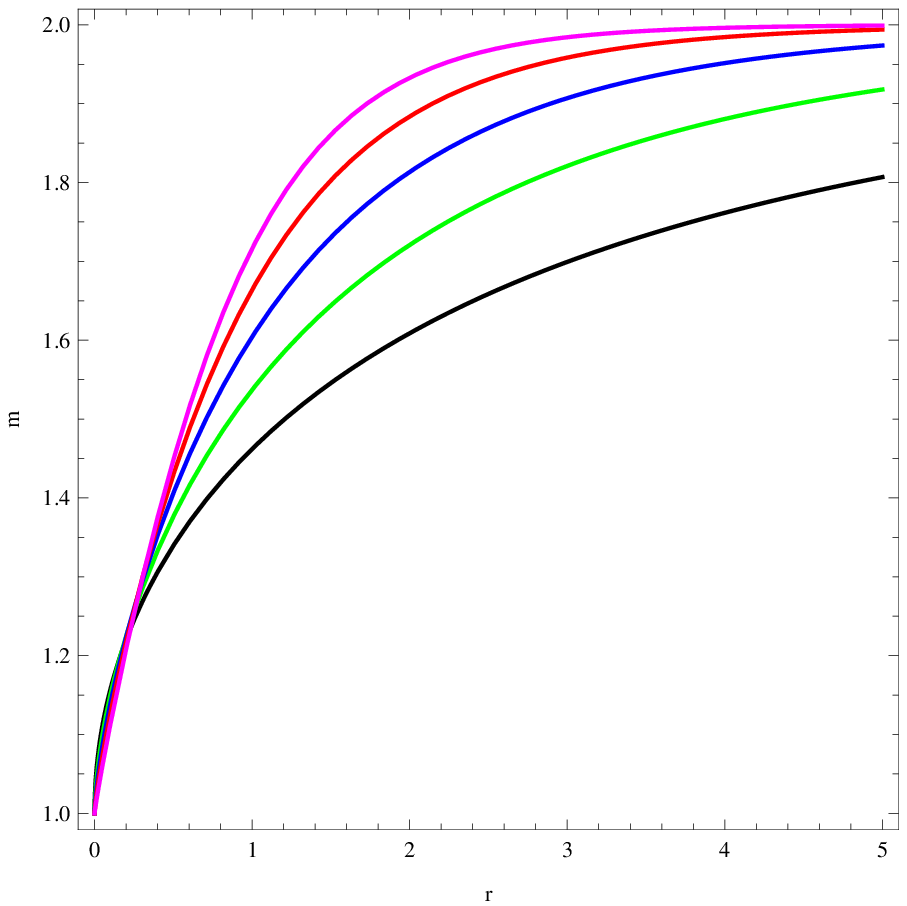}
\end{minipage} \hfill 
\begin{minipage}[t]{0.3\linewidth}
\includegraphics[width=\linewidth]{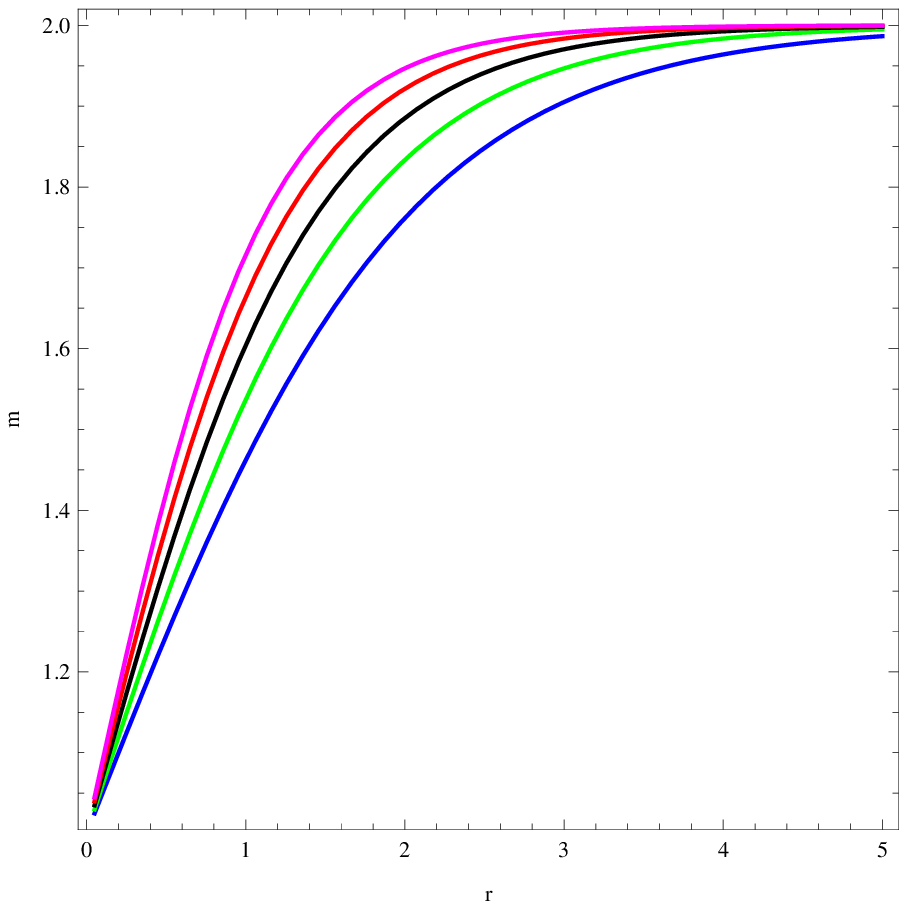}
\end{minipage} \hfill
\caption{{\protect\footnotesize These figures show the evolution of the mass as 
the radius evolves for the case $n<0$, for instance $n=-0.5$  (left diagram); 
for the case $0<n<1$, for instance $n=0.5$ (middle diagram) and for the case 
$n>1$, for instance $n=2$ (right diagram). For the three diagrams, the 
representative of the $TT$ are the black ones $(b_1=0)$. The red, magenta, blue 
and green curves are plotted for $b_1= -2; -1; 1; 2 $, respectively. The graphs 
here are related to neutron stars within polytropic EoS for power-law 
correction to $TT$.}}
\label{}
\end{figure}
\end{center}

\begin{center}
\begin{figure}[t]
\begin{minipage}[t]{0.3\linewidth}
\includegraphics[width=\linewidth]{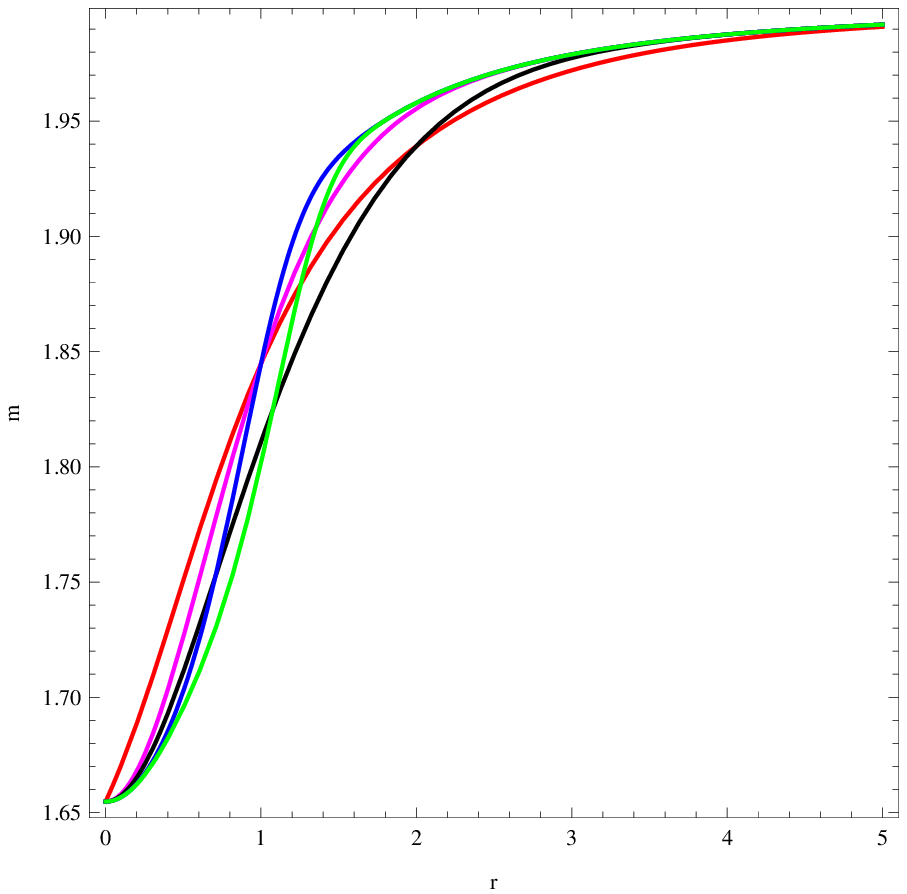}
\end{minipage} \hfill 
\begin{minipage}[t]{0.3\linewidth}
\includegraphics[width=\linewidth]{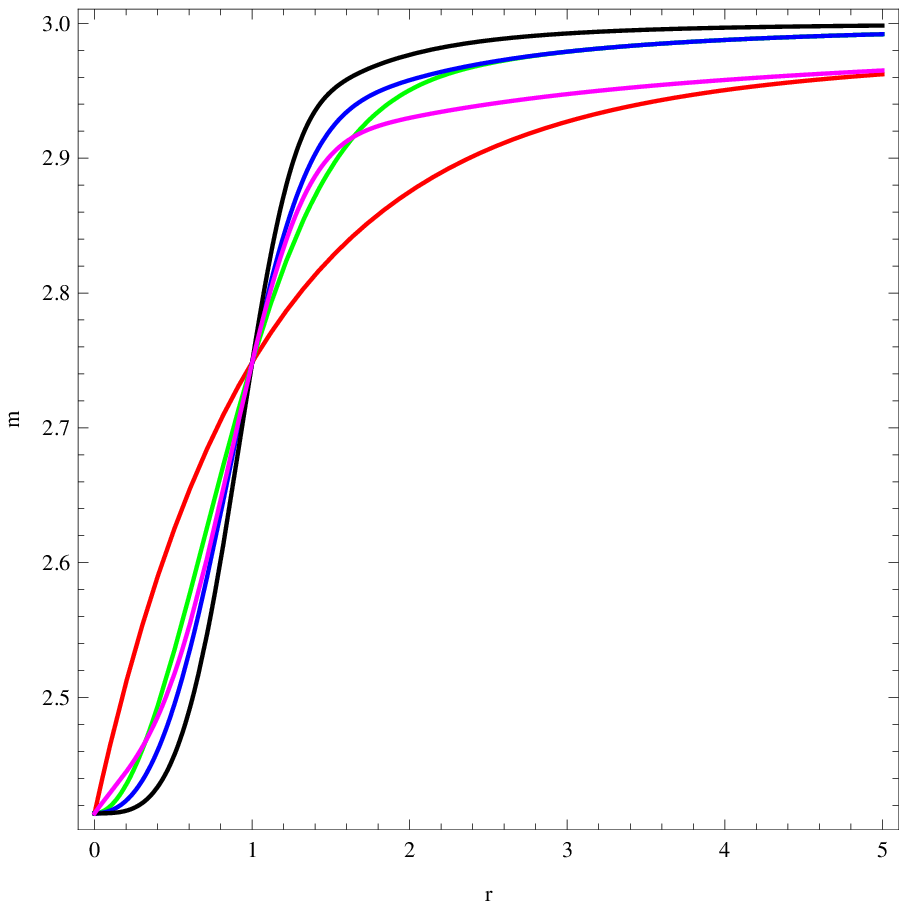}
\end{minipage} \hfill
\caption{{\protect\footnotesize The figures show the evolution of the mass as 
the radius evolves for the case $q<0$, for instance $q=-1$  (left diagram) and 
for the case $q>0$, for instance $q=1$ (right diagram). For both diagrams, the 
representative of the $TT$ are the black ones $(b_2=0)$. The red, magenta, blue 
and green curves are plotted for $b_2= -1; -0.5; 0.5; 1 $, respectively. 
The graphs here are related to neutron stars within polytropic EoS for 
exponential correction to $TT$.}}
\label{}
\end{figure}

\end{center}

\begin{center}
2.  {\it SLy EoS}
\end{center}
This type of equation of state characterizes the behaviour of nuclear matter at 
high densities and is expressed as
\begin{eqnarray}\label{fla26}
\zeta=\frac{a_1+a_2\xi+a_3\xi^3}{1+a_4\xi}f_0a_5\left(\xi-a_6\right)+f_0
\left(a_7+a_8\xi\right)\left[a_9\left(a_{10}-\xi\right)\right]\nonumber\\
+f_0\left(a_{11}+a_{12}\xi\right)\left[a_{13}\left(a_{14}-\xi\right)\right]+
f_0\left(a_{15}+a_{16}\xi\right)\left[a_{17}\left(a_{18}-\xi\right)\right].
\end{eqnarray}
Here the parameters $\zeta$ and $\xi$ are defined as in (\ref{fla25}) and 
the function $f_0$ is defined by
\begin{eqnarray}
f_0(x)=\frac{1}{e^x+1}
\end{eqnarray}
The values of coefficients $a_i$ can be view in the Table $1$ 
\cite{49deasuivre}.\par 
\vspace{0.25cm}
\begin{center}
\begin{tabular}{|c|c|c|c|}
\multicolumn{4}{c}{Table 1}\\
\hline i& $a_i$(SLy)& i& $a_i$(SLy)\\
\hline 1& 6.22 & 10 & 11.4950\\
\hline 2& 6.121 & 11 & -22.775\\
\hline 3& 0.005925 & 12 & 1.5707\\
\hline 4& 0.16326 & 13 & 4.3\\
\hline 5& 6.48 & 14 & 14.08\\
\hline 6& 11.4971 & 15 & 27.80\\
\hline 7& 19.105 & 16 & -1.653\\
\hline 8& 0.8938 & 17 & 1.50\\
\hline 9& 6.54 & 18 & \multicolumn{1}{|l|}{14.67}\\
\hline
\end{tabular}
\end{center}
As in the previous case, we assume that the density at the center of star 
evolves from a critical $\rho_c$ to the total density at it surface. The 
deviation from the $TT$ is  presented for the both power-law gravity and the 
exponential one for different values of the parameters $b_1$ and $b_2$. The 
evolution of the mass-radius diagram for neutron stars within SLy EoS are 
represented in Fig $3$ for power-law correction and Fig $4$ for exponential 
correction to the TT of gravity.


\begin{center}
\begin{figure}[t]
\begin{minipage}[t]{0.3\linewidth}
\includegraphics[width=\linewidth]{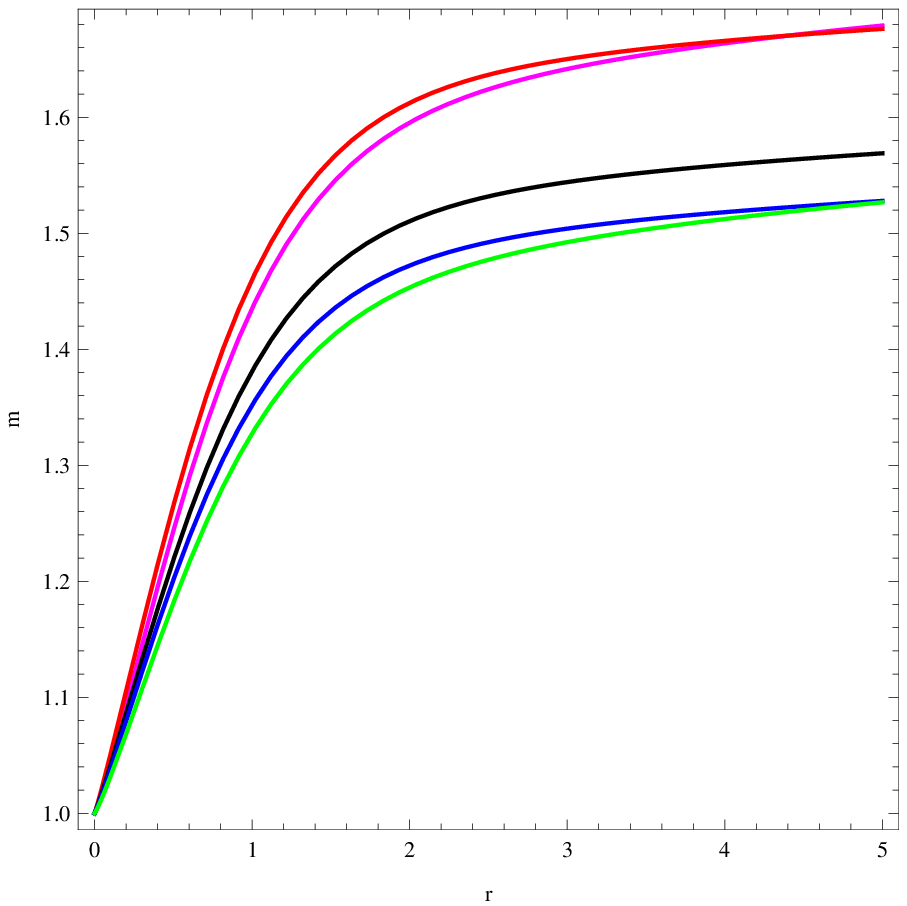}
\end{minipage} \hfill 
\begin{minipage}[t]{0.3\linewidth}
\includegraphics[width=\linewidth]{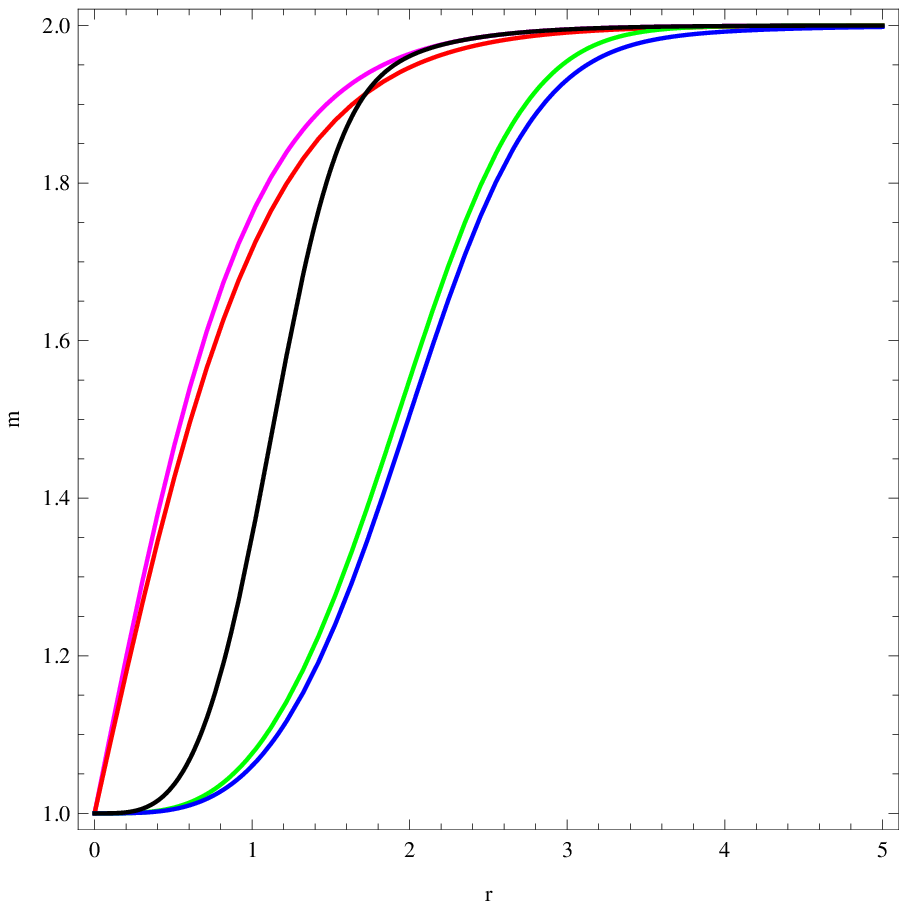}
\end{minipage} \hfill 
\begin{minipage}[t]{0.3\linewidth}
\includegraphics[width=\linewidth]{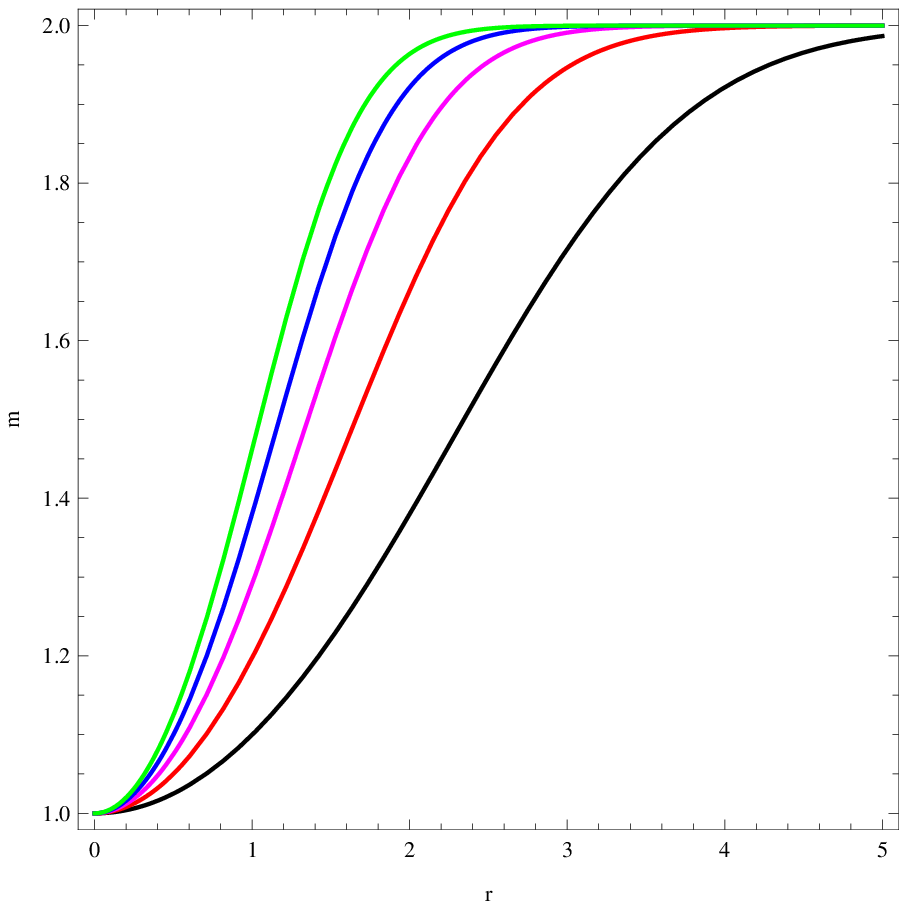}
\end{minipage} \hfill
\caption{{\protect\footnotesize These figures show the evolution of the mass as 
the radius evolves for the case $n<0$, for instance $n=-0.5$  (left diagram); 
for the case $0<n<1$, for instance $n=0.5$ (middle diagram) and for the case 
$n>1$, for instance $n=2$ (right diagram). For the three diagrams, the 
representatives  of the $TT$ are the black ones $(b_1=0)$. The red, magenta, 
blue and green curves are plotted for $b_1= -2; -1; 1; 2 $, respectively. The 
graphs here are related to neutron stars within SLy EoS for power-law 
correction to $TT$.}}
\label{}
\end{figure}
\end{center}

\begin{center}
\begin{figure}[hbtp]
\begin{minipage}[t]{0.3\linewidth}
\includegraphics[width=\linewidth]{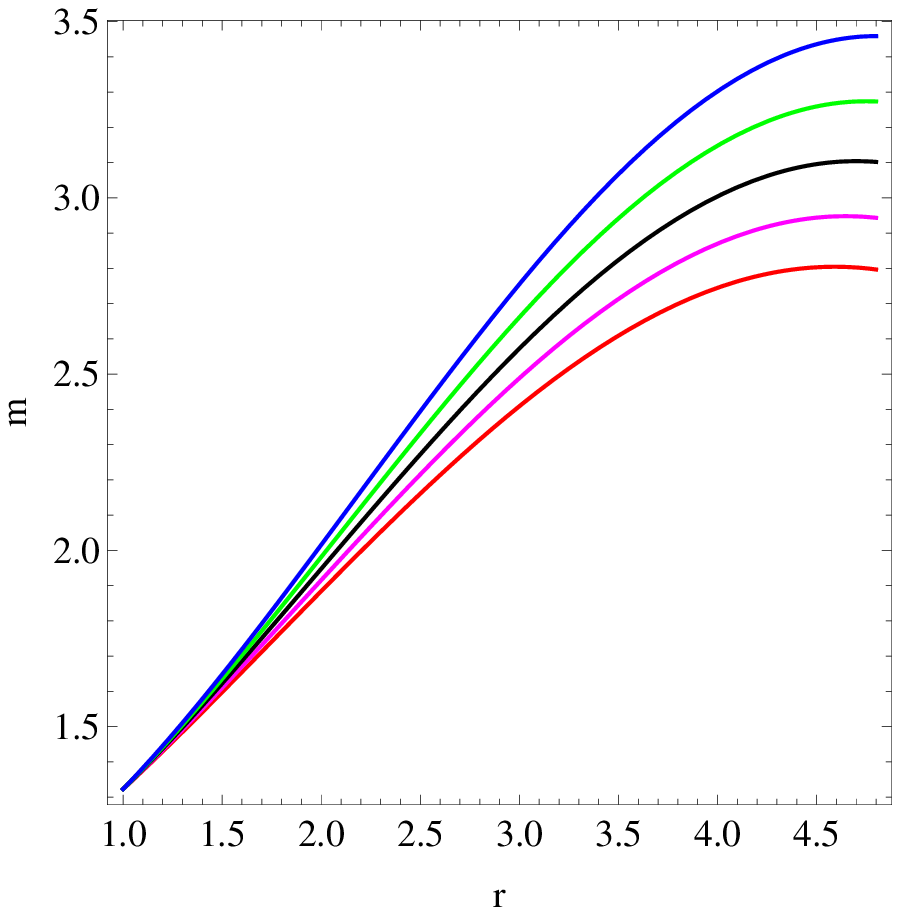}
\end{minipage} \hfill 
\begin{minipage}[t]{0.3\linewidth}
\includegraphics[width=\linewidth]{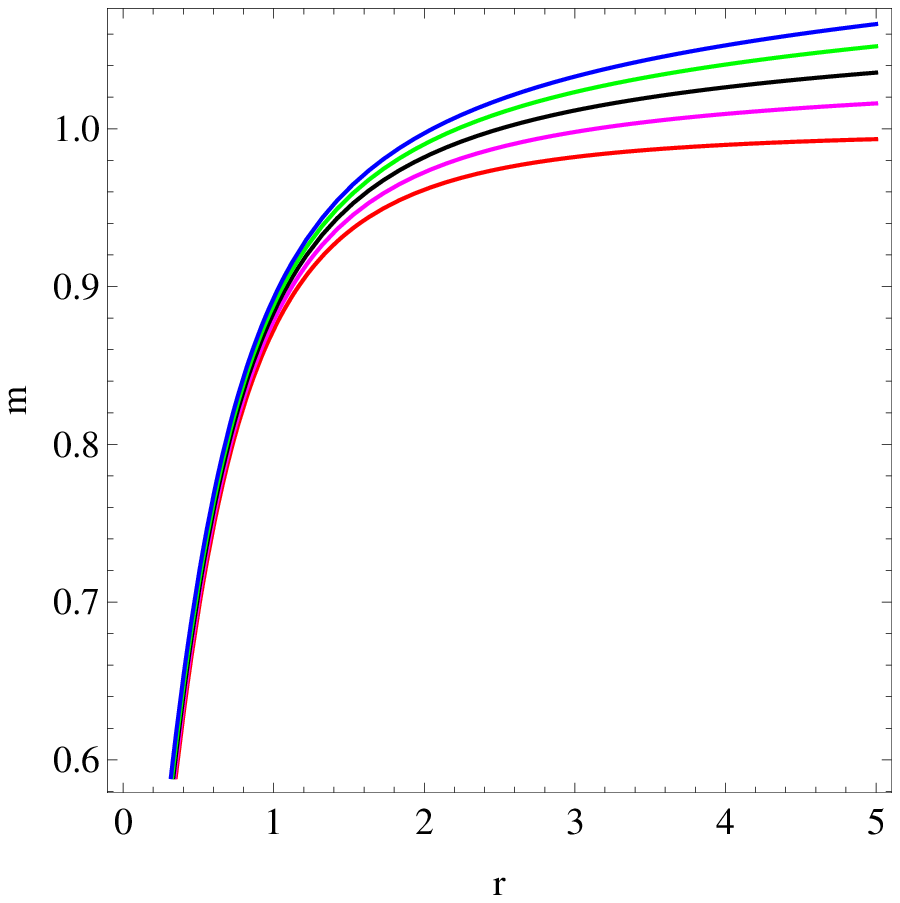}
\end{minipage} \hfill
\caption{{\protect\footnotesize The figures show the evolution of the mass as 
the radius evolves for the case $q<0$, for instance $q=-1$  (left diagram) and 
for the case $q>0$, for instance $q=1$ (right diagram). For both diagrams, the 
representatives of the $TT$ are the black ones $(b_2=0)$. The red, magenta, 
blue and green curves are plotted for $b_2= -1; -0.5; 0.5; 1 $, respectively. 
The graphs here are related to neutron stars within SLy EoS for exponential 
correction to $TT$.}}
\label{}
\end{figure}
\end{center}

\subsubsection{Quark Stars}
A quark star is a self-gravitating system consisting of deconfined $u$, $d$ and 
$s$ quarks and electrons \cite{30deodintsov}. These deconfined quarks are the 
fundamental elements of the color superconductor system. In the comparison with 
the standard hadron matter, they lead to a softer equation of state. In the 
frame of the so-called bag model, a simple equation of state is obtained for 
quark matter \cite{odintsov}
\begin{eqnarray}\label{fla18}
p=\omega(\rho-4\gamma)
\end{eqnarray}
where $\gamma$ is the bag constant. The value of the parameter $\omega$ depends
on the mass $m_s$ of the strang quark. For the radiation, one has $m_s=0$ and 
the parameter is $\omega=0$, and  for more realistic model where $m_s=250 MeV$, 
the parameter is $\omega=0.28$. The parameter $\gamma$ belongs to the interval 
$58.8 <B<91.2$ in the unit of $MeV/fm^3$ \cite{4deodintsov}. The evolution the 
mass-radius diagram for quark stars are represented in Fig $5$ for power-law 
correction and Fig $6$ for exponential correction to the TT of gravity.


\begin{center}
\begin{figure}[hbtp]
\begin{minipage}[t]{0.3\linewidth}
\includegraphics[width=\linewidth]{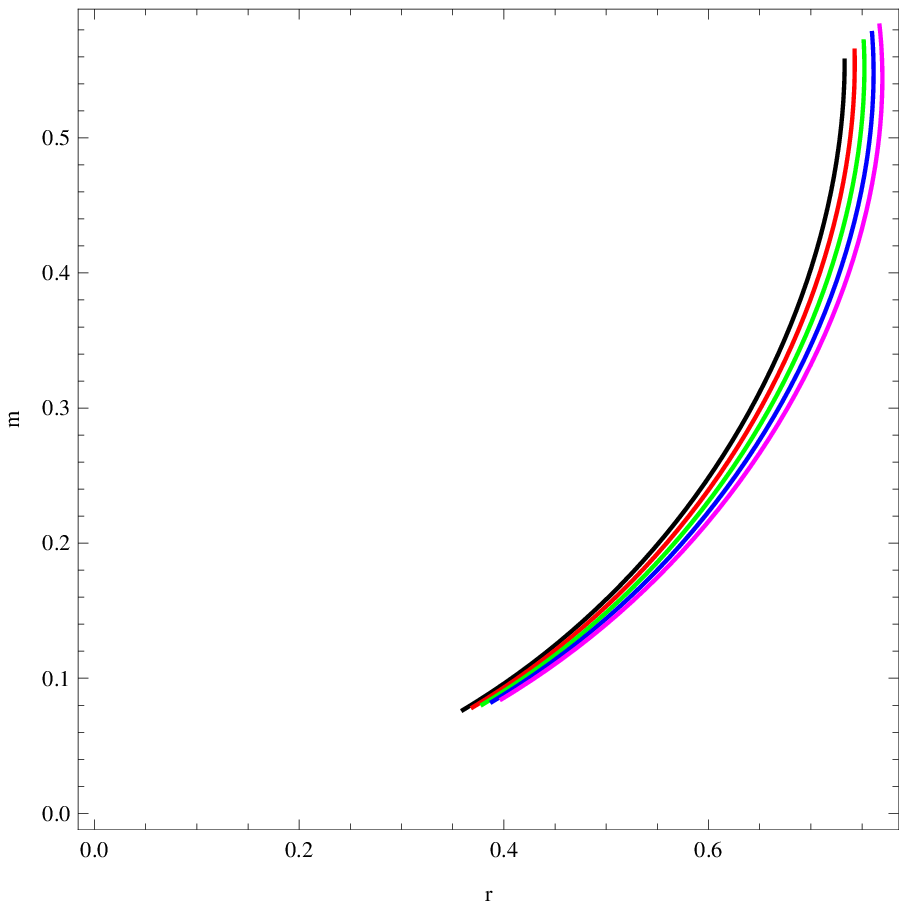}
\end{minipage} \hfill 
\begin{minipage}[t]{0.3\linewidth}
\includegraphics[width=\linewidth]{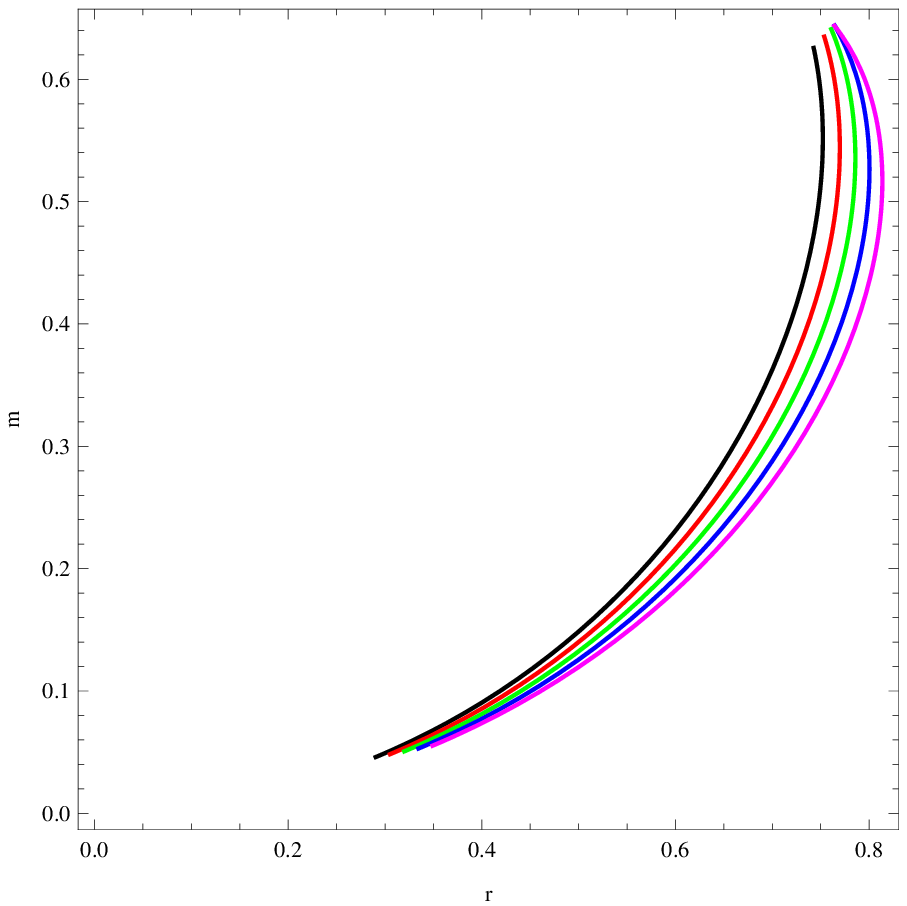}
\end{minipage} \hfill 
\begin{minipage}[t]{0.3\linewidth}
\includegraphics[width=\linewidth]{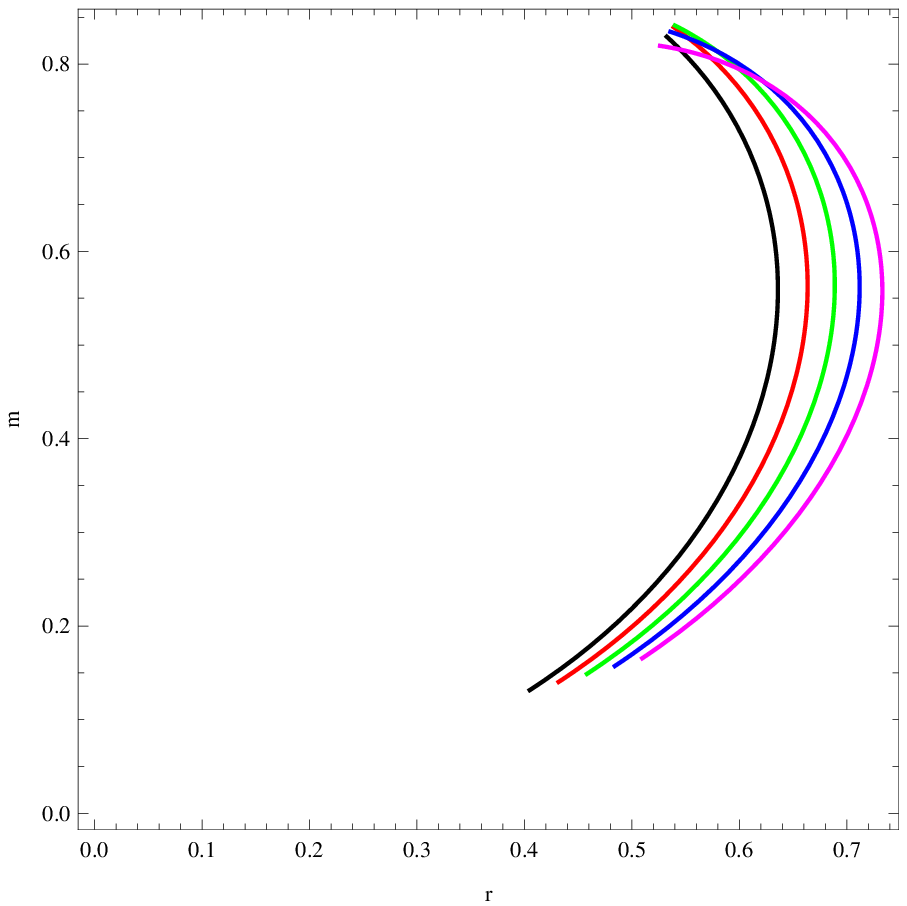}
\end{minipage} \hfill
\caption{{\protect\footnotesize These figures show the evolution of the mass as 
the radius evolves for the case $n<0$, for instance $n=-0.5$  (left diagram); 
for the case $0<n<1$, for instance $n=0.5$ (middle diagram) and for the case 
$n>1$, for instance $n=2$ (right diagram). For the three diagrams, the 
representatives of the $TT$ are the black ones $(b_1=0)$. The red, magenta, 
blue and green curves are plotted for $b_1= -2; -1; 1; 2 $, respectively. The 
graphs are related to the quark stars with the polytropic correction to TT.}}
\label{}
\end{figure}
\end{center}

\begin{center}
\begin{figure}[hbtp]
\begin{minipage}[t]{0.3\linewidth}
\includegraphics[width=\linewidth]{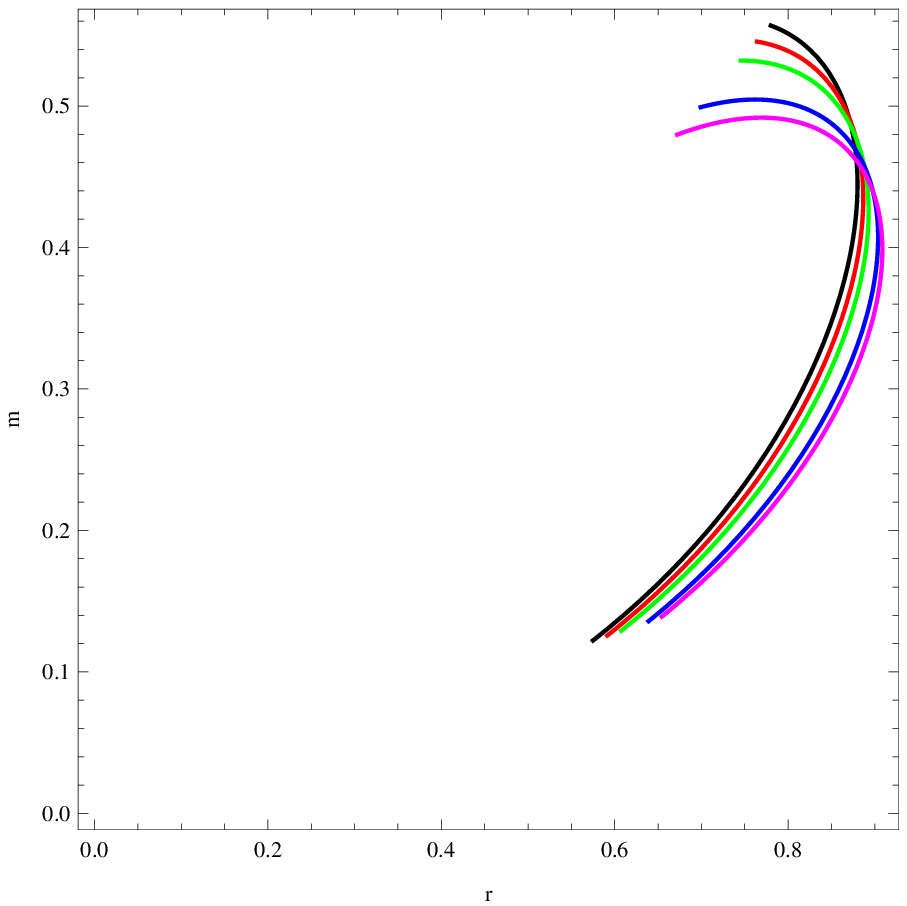}
\end{minipage} \hfill 
\begin{minipage}[t]{0.3\linewidth}
\includegraphics[width=\linewidth]{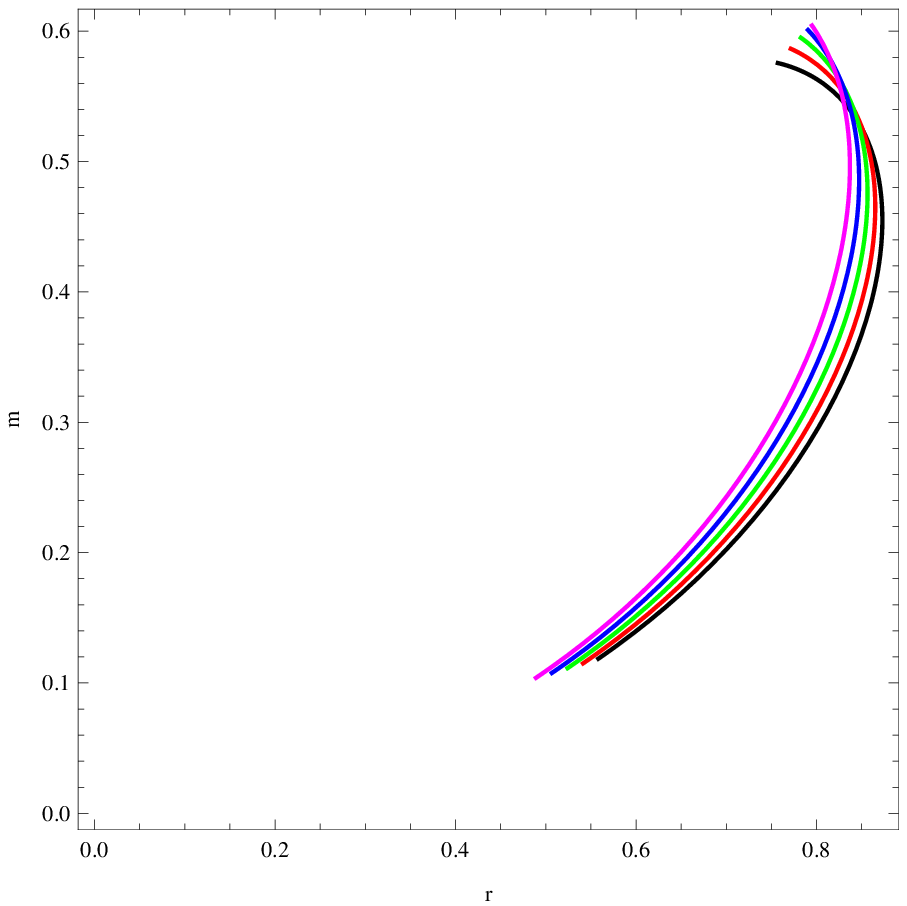}
\end{minipage} \hfill
\caption{{\protect\footnotesize The figures show the evolution of the mass as 
the radius evolves for the case $q<0$, for instance $q=-1$  (left diagram) and 
for the case $q>0$, for instance $q=1$ (right diagram). For both diagrams, 
the representatives of the TT are the black ones ($b_2=0$). The red, magenta, 
blue and green curves are plotted for $b_2= -1; -0.5; 0.5; 1 $, respectively. 
The graphs here are related to neutron stars within SLy EoS for exponential 
correction to TT.}}
\label{}
\end{figure}
\end{center}

\section{Conclusion} 
The structures of neutron and quark stars structures have been analyzed in this 
work in the framework of $f(T)$, $T$ being the torsion scalar. The main goal 
here is searching the deviation of mass-radium diagram of both neutron and quark
stars of the modified gravity, view as correction terms to the $TT$, from 
corresponding diagram of the $TT$. The first has been establishing the TOV 
equations for $f(T)$ theory. The fundamental fashion here is assuming the 
non-diagonal tetrad from which any constraint can occur about the choice of the 
algebraic action function. Therefore, we assume two interesting cases as 
correction terms; the power-law and exponential terms, including the parameters 
($b_1$ and $n$) and  ($b_2$ and $q$), respectively. \par 
Our analysis are based on the numerical integration of the system formed by TOV 
equations and the different EoS, the polytropic and SLy EoS for neutron stars 
and a suitable EoS for the quark stars. Our results show that for some values 
of the input parameters in the neutron stars case, for both polytropic and SLy 
EoS, the deviation of the $f(T)$ terms from the TT is obvious for the beginning 
and the intermediate values of the radius. However, in some cases and for large 
values of the radius, the correction terms do not have effect on the evolution 
of the mass, with respect to the $TT$ case. In the quark stars case, it appears 
that for any value of the radius, the deviation of the $f(T)$ theory terms from the $TT$ is obvious. 

\vspace{1cm}

{\bf Acknowledgments:} A.V.Kpadonou and M.J.S.Houndjo thank {\it Ecole Normale 
Sup\'erieure de Natitingou} for partial financial support.
M.E.Rodrigues expresses his sincere gratitude do UFPA and CNPq for partial 
financial support during the elaboration of this work.

\newpage


\end{document}